\documentstyle[prc,aps,preprint]{revtex}
\begin{document}
\draft
\date{\today}
\title{$\omega$-Meson
Photoproduction on Nucleons In the Near Threshold Region}
\author{H. Babacan, T. Babacan, A. Gokalp, O. Yilmaz}
\address{Physics Department, Middle East Technical University, 06531
Ankara, Turkey}
\maketitle
\begin{abstract}

The $\omega$-meson photoproduction, $\gamma+p\rightarrow
p+\omega$, is studied in the framework of a model, containing
$\pi$-meson exchange in t-channel and nucleon-exchange in
s- and u-channels. Considering both $\omega NN$-coupling
constants in the region of time-like meson four momenta as the
free parameters, we find different sets of solutions for these
constants from the existing data on the t-dependence of the
differential cross sections, $d\sigma(\gamma+p\rightarrow
p+\omega)/dt$, in the near  threshold region $E_{\gamma}\leq 2$
GeV. These sets of $\omega NN$-coupling constants,
corresponding to destructive and constructive $\pi\bigotimes
N$-interference contributions to $d\sigma/dt$ can be well
distinguished by measurements of beam asymmetry, induced by linear
photon polarization.
\end{abstract}

\section{Introduction}

The vector meson photoproduction on nucleons, $\gamma+p\rightarrow
p+V$, $V=\rho$ or $\omega$, in the near threshold region
$E_{\gamma}< 2$ GeV, can be considered as a source of
important information concerning interesting problems of hadron
electrodynamics, such as for example the values of different
electromagnetic and strong coupling constants and the properties of
the so-called "missing" resonances \cite{R1,R2}. To solve these and
other similar problems a suitable model for $\gamma+p\rightarrow p+V$
must be formulated. This is especially important for the study of
the physics of missing resonances. It is a well known fact that in the
$N^{*}$-resonance physics for the successful extraction of
adequate resonance information, the correct theoretical description
of nonresonant mechanisms must be at hand. This is not a simple task and
it has been an actual problem up to now even for the case of "oldest"
$\Delta(1232)$-resonance, where for the exact value of the small
quadrupole (E2) multipole amplitude for the decay $\Delta\rightarrow
N+\gamma$, the correct knowledge of the corresponding
nonresonant background is needed \cite{R3,R4,R5}. Evidently, this
statement is correct for any $N^{*}$-resonance.

The theoretical study of the nonresonant mechanisms for the
processes of vector meson photoproduction on nucleons,
$\gamma+N\rightarrow N+V$, in the near threshold region
$E_{\gamma}< 2$ GeV, is at its beginnings, there is no unique and
well-proved solution of this task. The following mechanisms are
considered in the literature
\cite{R1,R2,R6,R7,R8,R9,R10,R11,R12,R13}: pseudoscalar ($\pi, \eta$) and
scalar ($\sigma$)-meson exchanges in t-channel, one-nucleon
exchanges in (s+u)-channels, and Pomeron exchange for the case of
neutral vector meson photoproduction. Typically, different
combinations of these contributions are analyzed by different
authors. All these ingredients are characterized by relatively
large number of coupling constants and cut-off parameters which
determine the phenomenological form factors for the electromagnetic
and strong vertexes of the considered pole diagrams. Some of these
parameters can be determined from other processes, such as for
example, the radiative decays of vector mesons
$V\rightarrow\pi(\eta)+\gamma$, with good enough accuracy. The
same is correct for the $\pi NN$-coupling constant, which has been
determined with the highest accuracy among  different strong
coupling constants. However, another situation exits for VNN-coupling
constants which determine the nucleon pole diagrams for processes
$\gamma+N\rightarrow N+V$ in the region of time-like momentum of
vector meson, $q^{2}= m_{v}^{2}$. In general, therefore,
these values can be very different from their values in the space-like
region of vector meson momentum, which is the case of pion
photoproduction,
$\gamma+N\rightarrow N+\pi$, vector meson exchange in t-channel, or
NN-potential \cite{R14,R15}.

Another important question concerns applicability of Pomeron
exchange in the near threshold region, where the validity of the
Regge-regime seems  problematic, and not so evident. It is
enough to remember that, typically the kinematic region of
application of Regge-theory  is determined by the following
conditions: $s\gg M^{2}, s\gg t$, where s and t are the standard
Mandelstam variables, M is the nucleon mass. Evidently such
conditions can not be realized in the near threshold region for
$\gamma+N\rightarrow N+V$.

In this work, we attempt to consider these questions for the
$\omega$-meson photoproduction, $\gamma+N\rightarrow N+\omega$,
which have some special properties, different from
$\rho^{0}$-meson photoproduction, $\gamma+N\rightarrow
N+\rho^{0}$. First of all, due to the relatively large
$\omega\pi\gamma$-coupling constant in comparison with
$\rho\pi\gamma$-coupling constant, one-pion contribution can be
considered as
the main mechanism in the near threshold region for
$\gamma+N\rightarrow N+\omega$ processes. That is an important point
because
this contribution is determined by product of the well-known
coupling constants, $g_{\omega\pi\gamma}g_{\pi NN}$. So here we
have a situation, different from the case of
$\rho^{0}$-photoproduction, where another t-exchange is important,
namely $\sigma$-exchange. But properties of the
$\sigma$-meson are not well established now, even its mass is
inside of a wide interval: 400-1200 MeV \cite{R16}, the same is also true
for the $\sigma$-width: $\Gamma=600-1000$ MeV. Moreover, the product of
necessary coupling constants, namely $g_{\rho\sigma\gamma}g_{\sigma
NN}$ cannot be considered as well known. For example, the
"standard" assumption \cite{R13} that $\rho\sigma\gamma$-coupling
constant is essentially larger in comparison with
$\omega\sigma\gamma$-coupling constant must be revised now after
the experiment of Novosibirsk group \cite{R17}, which proved
definitely that the width of radiative decays
$\omega\rightarrow\pi^{0}+\pi^{0}+\gamma$ and
$\rho\rightarrow\pi^{0}+\pi^{0}+\gamma$ are comparable. Let us note in
this connection the previous conclusion about large enough
$\rho\sigma\gamma$-coupling constant that was obtained on the
basis of the relatively large measured branching ratio for
$\rho\rightarrow\pi^{+}+\pi^{-}+\gamma$ in comparison with
$\omega\rightarrow\pi^{0}+\pi^{0}+\gamma$ branching ratio \cite{R18}.
However, the main contribution to
$\rho\rightarrow\pi^{+}+\pi^{-}+\gamma$ must be not due  to
$\sigma$-mechanism
($\rho^{0}\rightarrow\gamma+\sigma^{0}\rightarrow\gamma+\pi^{+}+
\pi^{-}$)
but due to $\gamma$-radiation of final charged pions.

Therefore, the situation with $\sigma$-exchange in the process
$\gamma+p\rightarrow p+\rho^{0}$ becomes more complicated now. In
principle it is possible to "save" $\sigma$-exchange in
$\gamma+p\rightarrow p+\rho^{0}$: the decrease in the value of
$g_{\rho\sigma\gamma}$-coupling constant , which follows from the
Novosibirsk experiment, can be compensated by correspondingly
increasing the value of $g_{\sigma NN}$ coupling constant. By such
a manipulation it is possible to conserve the substantial
$\sigma$-contribution to the matrix element for process $\gamma
+p\rightarrow p+\rho^{0}$ in the near threshold region, but as a
result, we will obtain quadratically increasing
$\sigma$-contribution to NN-potential. Therefore, this problem
must be studied independently.

And another point with increasing $\sigma NN$-coupling constant is
that this will also
increase respectively $\sigma$-contribution to the matrix element
of the process $\gamma
+p\rightarrow p+\omega$ making this contribution comparable with
that of the process $\gamma+p\rightarrow p+\rho^{0}$. So, in such
a situation,
we will have large and comparable contributions, namely $\sigma$
and $\pi$, to the matrix element of $\gamma+p\rightarrow p+\omega$ which
evidently contradicts the existing explanation of experimental
data about differential cross sections for this process. In order to
remove this contradiction, these two large contributions
must be essentially compensated by some destructive interference
with other possible contributions to the matrix element of
$\gamma+p\rightarrow p+\omega$ process. But pure imaginary
Pomeron contribution cannot interfere with real
amplitudes of $\pi$- and $\sigma$-exchange. So the best candidate for
such interference could be N-contribution considered in (s+u)-channels
to satisfy gauge invariance, or $N^{*}$-contributions.

We like to note that in the general case each $N^{*}$-resonance, with spin
$J\geq 3/2$, produces complicated enough spin structure in
the matrix element due to the possible six independent multipole
amplitudes, which must be nonzero.
In any case, the situation with resonance physics in
$\gamma+N\rightarrow N+V$ processes is evidently more complicated
in comparison with the pseudoscalar meson photoproduction on
nucleon: $\gamma+N\rightarrow N+\pi$ or $\gamma+N\rightarrow
N+\eta$. This means that the polarization phenomena in processes
$\gamma+N\rightarrow N+V$ are especially important to realize more
or less unique multipole analysis.

The specific property of $N^{*}$-contributions in s-channel is
the generation of the complex amplitudes. This new property of the
corresponding model will result in rich and specific T-odd
polarization effects in $\gamma+N\rightarrow N+V$,
such as the analyzing power induced by
the polarized nucleon target, or the polarization of produced
nucleon. So, namely the T-odd polarization phenomena in
$\gamma+N\rightarrow N+V$ will be the most decisive for the
estimation of $N^{*}$-contributions in a more definite way. Being the
simplest among all vector meson photoproduction processes,
the reaction $\gamma+N\rightarrow N+\omega$ seems as the most
suitable for the identification of the adequate nonresonant
mechanisms for such processes in the near threshold region.

In this paper, we try to estimate the role of nucleon
contribution to the matrix element of the processes
$\gamma+N\rightarrow N+\omega$. Instead of standard
and oversimplified model for $\gamma+p\rightarrow p+\omega$ with
$\pi$-exchange only we consider here more complicated
($\pi+N$)-model, but without Pomeron exchange in the near threshold
region. To estimate possible strong dependence of $\omega
NN$-coupling constants on vector meson four momenta, going from the region
of space-like
to time-like vector meson momentum, we consider in our approach
both possible $\omega NN$-coupling constants, tensor and vector types,
as free parameters to be determined by performing a fit to the
existing experimental data on the differential cross section
$d\sigma(\gamma+p\rightarrow p+\omega)/dt$ in the near threshold
region. Such a model will
produce non-trivial and relatively intensive polarization phenomena in
$\gamma+N\rightarrow N+\omega$. Of course, all these polarization
effects have  T-even character. But instead of trivial
polarization effects of the $\pi$-exchange model, the ($\pi+N$)-model will
produce specific t-behaviour of such observables, such as the asymmetry
$\Sigma$ induced by photon linear polarization,  the elements of
density matrix  for the vector mesons produced in collisions of
polarized and unpolarized particles. Among  the possible two spin
polarization observables of T-even nature let us note the
asymmetry in collisions of circularly polarized photons with
polarized targets. High energy photon beams with high degree of
circular polarization is available in JLAB now. Note also that the
suggested model with ($\pi+N$)-contributions
will produce also essential difference in observables on proton and
neutron targets due to $\pi\bigotimes N$-interference and due to
different N-contributions.

So our main aim in this work is to find a special simple ($\pi+N$)-model
with
relatively large N-contribution, which is cancelled in
differential cross section with unpolarized particles by the
essential $\pi\bigotimes N$-interference, as a result imitating the
differential cross section $d\sigma(\gamma+p\rightarrow p+\omega)/dt$ of
the pure $\pi$-exchange model. Evidently such ($\pi+N$)-model and
simple $\pi$-exchange model will differ essentially in isotopic effects
and
in polarization phenomena.

\section{Description of The Model}

We begin here by discussing the main properties of the suggested model for
the
process $\gamma+N\rightarrow N+\omega$ in the near threshold
region. The nucleon s-channel contribution is described by the
following amplitude:

\begin{eqnarray}
{\cal M}_{s}= \frac{e}{s-M^{2}} \overline{u}(p_{2})(g_{\omega
NN}^{V}\hat{U}+
\frac{g_{\omega NN}^{T}}{2 M }
\hat{U}~\hat{q})(\hat{p_{1}}+\hat{k}+M) (Q_{N}
\hat{\varepsilon}- \frac{\kappa_{N}}{2 M}
\hat{\varepsilon}~\hat{k}) u(p_{1})~~,\nonumber \\
\end{eqnarray}
where $\varepsilon_{\mu}$ and k ($U_{\mu}$ and q) are the
polarization four-vector and four-momentum of the photon
($\omega$-meson), $\varepsilon\cdot k= U\cdot
q=0$, $\hat{a}=\gamma^{\mu}a_{\mu}$, M is the
nucleon mass, $Q_{N}$ is the nucleon electric charge, i.e
$Q_{N}=1(0)$ for proton (neutron), $\kappa_{N}$ is the nucleonic
anomalous magnetic moment, $\kappa_{N}=1.79(-1.91)$ for proton
(neutron); $g_{\omega NN}^{V}$ and $g_{\omega NN}^{T}$ are the vector
(Dirac) and tensor
(Pauli) coupling constants of the $\omega NN$-vertex. The notation of
particle four momenta is presented in Fig. 1. We consider
here the quantities $g_{\omega NN}^{V}$ and $g_{\omega NN}^{T}$ as
constants,
neglecting their possible dependence on the virtuality s of the
intermediate nucleon. Therefore, the same coupling constants $g_{\omega
NN}^{V}$ and
$g_{\omega NN}^{T}$
determine the matrix element of nucleon exchange in u-channel as

\begin{eqnarray}
{\cal M}_{u}= \frac{e}{u-M^{2}} \overline{u}(p_{2})(Q_{N}
\hat{\varepsilon}- \frac{\kappa_{N}}{2 M}
\hat{\varepsilon}~\hat{k})(\hat{p_{2}}-\hat{k}+M) (g_{\omega
NN}^{V}\hat{U}+
\frac{g_{\omega NN}^{T}}{2 M } \hat{U}~\hat{q}) u(p_{1})~~,\nonumber \\
\end{eqnarray}
where $u=(k-p_{2})^{2}$.

We like to repeat here once more that, in the general case the
quantities  $g_{\omega NN}^{V}$ and $g_{\omega NN}^{T}$ in Eq.(2) must be
considered as
some form factors, $g_{i}=g_{i}(u)$, but to preserve gauge
invariance of the sum ${\cal M}_{s}+{\cal M}_{u}$, we will neglect
the possible s- or u-dependence of $g_{i}$. In any case we consider
here the very probable possibility that
the $g_{\omega NN}^{V}$ and $g_{\omega NN}^{T}$ in Eqs.(1) and (2) are
different from their
values in the space-like region of vector meson momentum, i.e. we will
consider the coupling
constants $g_{\omega NN}^{V}$ and $g_{\omega NN}^{T}$ as the fitting
parameters of the suggested model.

The matrix element for t-channel $\pi$-meson exchange can be
written straightforwardly in the following way:

\begin{equation}
 {\cal M}_{t}
 =e~\frac{g_{\omega\pi\gamma}}{m_{\omega}}~\frac{g_{\pi NN}}{t-m_{\pi}^{2}}
 F_{\pi NN}(t)~F_{\omega\pi\gamma}(t)~(\overline{u}(p_{2})~\gamma_{5}~u(p_{1}))
 ~(\epsilon^{\mu\nu\alpha\beta}
 ~\varepsilon_{\mu}~k_{\nu}~U_{\alpha}~q_{\beta})~,\nonumber \\ \nonumber
\\
\end{equation}
where $t=(k-q)^{2}$, $m_{\pi}$ is the pion mass, $m_{\omega}$ is
the $\omega$-meson mass, $g_{\omega\pi\gamma}$ and
$F_{\omega\pi\gamma}(t)$ ($g_{\pi NN}$ and $F_{\pi NN}(t)$) are
the coupling constant and the corresponding form factor for the
electromagnetic-$\omega\pi\gamma$ (strong-$\pi NN$) vertex of the
considered diagram.

We like to note that, in our analysis we avoid using any form factor in
${\cal M}_{s}+{\cal M}_{u}$, again to preserve  gauge
invariance of ${\cal M}_{s}+{\cal M}_{u}$. Evidently, a
s-dependent form factor for ${\cal M}_{s}$ and  a u-dependent form
factor for ${\cal M}_{u}$, which seems as the most natural way to
introduce form factors,
will destroy the coherence of both of these contributions with respect
to the conservation of the electromagnetic current for the
considered process.

We prefer in our treatment to include the possible form factor effects and
also the effects of transition from space-like to time-like region in
$\omega$-meson four momentum in the effective
values of the coupling constants $g_{\omega NN}^{V}$ and $g_{\omega
NN}^{T}$. If the above
mentioned effects are important, the resulting values of fitted
coupling constants $g_{\omega NN}^{V}$ and $g_{\omega NN}^{T}$, which are
to be obtained by a fit
to the existing experimental data about differential cross section for
$\gamma+p\rightarrow p+\omega$ \cite{R19}, will be different from
their values obtained in the space-like region. Therefore, these
new
values for the coupling constants $g_{\omega NN}^{V}$ and $g_{\omega
NN}^{T}$ can also be
used in similar analysis of the nucleon contribution to many other
processes with $\omega$-meson production, such as, $\pi+N\rightarrow
N+\omega$, $e^{-}+N\rightarrow e^{-}+N+\omega$, $\pi+N\rightarrow
\pi+\omega$ etc.

In our calculation of different observables for
$\gamma+N\rightarrow N+\omega$ we use the formalism of so
called transversal amplitudes in the center of mass system (CMS) of the
considered reaction.
This formalism is effective for the analysis of polarization
phenomena in the processes of the vector meson photoproduction,
and especially useful in the analysis of the problem of the full
reconstruction of the spin structure of the matrix element for
$\gamma+N\rightarrow N+V$ from the
complete experimental data.

The corresponding parametrization of the general matrix element of
any photoproduction process  $\gamma+N\rightarrow N+V$, which is valid for
any
model, can be written in terms of 12 independent transversal spin
structures in the following way:
\begin{eqnarray}
 {\cal M}& = & \varphi_{2}^{\dag}{\cal F}\varphi_{1}~~, \nonumber \\
 {\cal F}& = & if_{1}(\vec{\varepsilon}.\hat{\vec{m}})(\vec{U}.\hat{\vec{m}})
 + if_{2}(\vec{\varepsilon}.\hat{\vec{m}})(\vec{U}.\hat{\vec{k}})
 + if_{3}(\vec{\varepsilon}.\hat{\vec{n}})(\vec{U}.\hat{\vec{n}})\nonumber \\
 &+&(\vec{\sigma}.\hat{\vec{n}})[f_{4}(\vec{\varepsilon}.\hat{\vec{m}})(\vec{U}.\hat{\vec{m}})
 + if_{5}(\vec{\varepsilon}.\hat{\vec{m}})(\vec{U}.\hat{\vec{k}}) +
 if_{6}(\vec{\varepsilon}.\hat{\vec{n}})(\vec{U}.\hat{\vec{n}})]\nonumber \\
 &+&(\vec{\sigma}.\hat{\vec{m}})[f_{7}(\vec{\varepsilon}.\hat{\vec{m}})(\vec{U}.\hat{\vec{n}})
 + if_{8}(\vec{\varepsilon}.\hat{\vec{n}})(\vec{U}.\hat{\vec{m}}) +
 if_{9}(\vec{\varepsilon}.\hat{\vec{n}})(\vec{U}.\hat{\vec{k}})]\nonumber \\
 &+&(\vec{\sigma}.\hat{\vec{k}})[f_{10}(\vec{\varepsilon}.\hat{\vec{m}})(\vec{U}.\hat{\vec{n}})
 + if_{11}(\vec{\varepsilon}.\hat{\vec{n}})(\vec{U}.\hat{\vec{m}})
 + if_{12}(\vec{\varepsilon}.\hat{\vec{n}})(\vec{U}.\hat{\vec{k}})]~~,
\end{eqnarray}
where the set of unit orthogonal 3-vectors $\hat{\vec{m}}$,
$\hat{\vec{n}}$, and $\hat{\vec{k}}$ are defined as: $
\hat{\vec{k}}=\vec{k}/|\vec{k}|,
\hat{\vec{n}}=\vec{k}\times\vec{q}/|\vec{k}\times\vec{q}|,
\hat{\vec{m}}= \hat{\vec{n}}\times \hat{\vec{k}}$ and $\vec{k}$
and  $\vec{q}$ are the three-momentum of the photon and the vector meson
in CMS,
$\varphi_{1}$ and $\varphi_{2}$ are the two-component spinors for
initial and final nucleons;  $f_{i}$, $i=1,...,12$, are the
so-called transversal amplitudes, which are complex functions of
two independent invariant variables, s and t, $f_{i}=f_{i}(s,t)$.

The differential cross section $d\sigma/d\Omega$ with all the particles
in the initial and final states unpolarized, and
the beam asymmetry $\Sigma$ which is defined as

\begin{equation}
 \Sigma=\frac{d\sigma_{\parallel}/d\Omega-d\sigma_{\perp}/d\Omega}{d\sigma_
 {\parallel}/d\Omega+d\sigma_{\perp}/d\Omega}~~,
\end{equation}
can be expressed as the following quadratic combinations of the
transversal amplitudes $f_{i}$:
\begin{eqnarray}
 \frac{d\sigma}{d\Omega}& = &{\cal N}\left(h_{1}+h_{2}\right)~~,\nonumber \\
 \Sigma& = & \frac{\left(h_{1}-h_{2}\right)}{\left(h_{1}+h_{2}\right)}~~,
 \nonumber \\
 h_{1}&=& \frac{1}{2}\{\left[|f_{1}|^{2}+|f_{2}|^{2}+|f_{4}|^{2}+|f_{5}|^{2}
 +|f_{7}|^{2} +|f_{10}|^{2}\right]\nonumber \\
 &+&\left[\frac{q^{2}\sin^{2}\theta}{m_{v}^{2}}\right]\left[|f_{1}|^{2}
 +|f_{4}|^{2}\right]+\left[\frac{q^{2}\cos^{2}\theta}{m_{v}^{2}}\right]
 \left[|f_{2}|^{2}+|f_{5}|^{2}\right]\nonumber\\
 &+&\left[\frac{q^{2}2\sin\theta\cos\theta}{m_{v}^{2}}\right]Re\left[(f_{1}
 f_{2}^{*})+(f_{4}f_{5}^{*})
 \right]\}~~,\nonumber\\
 h_{2} &=& \frac{1}{2}\{\left[|f_{2}|^{2}+|f_{6}|^{2}+|f_{8}|^{2}+|f_{9}|^{2}
 +|f_{11}|^{2}+|f_{12}|^{2}\right]\nonumber\\
 &+&\left[\frac{q^{2}\sin^{2}\theta}{m_{v}^{2}}\right]\left[|f_{8}|^{2}
 +|f_{11}|^{2}\right]+
 \left[\frac{q^{2}\cos^{2}\theta}{m_{v}^{2}}\right]\left[|f_{9}|^{2}+|f_{12}|
 ^{2}\right]\nonumber \\
 &+&\left[\frac{q^{2}2\sin\theta\cos\theta}{m_{v}^{2}}\right]Re\left[(f_{8}
 f_{9}^{*})+(f_{11}f_{12}^{*})
 \right]\}~~,
\end{eqnarray}
where ${\cal N}=|\vec{q}|/64\pi^{2} s |\vec{k}|$ and
$d\sigma_{\parallel}/d\Omega$ ($d\sigma_{\perp}/d\Omega$) is the
cross section of photon absorption with linear polarization which
is parallel (orthogonal) to the reaction plane.

Let us give now, as an example, the expressions for $f_{i}$
corresponding to $\pi$- and  $\sigma$-exchange:

\begin{eqnarray}
 f_{i\pi} & = & e~~\frac{g_{\omega\pi\gamma}}{m_{\omega}}~~
 \frac{g_{\pi NN}}{t-m_{\pi}^{2}}~~\sqrt{(E_{1}+M)~(E_{2}+M)}
 ~~f^{\prime}_{i\pi}~~,\nonumber \\
 f^{\prime}_{1\pi}& = &~~f^{\prime}_{2\pi}~~=~~f^{\prime}_{3\pi}~~=~
 ~f^{\prime}_{4\pi}~~=~~f^{\prime}_{5\pi}~~=~~f^{\prime}_{6\pi}~~=
 ~~0\nonumber \\
 f^{\prime}_{7\pi}& = & \frac{|\vec{q}|}{E_{2}+M}~~B_{1\pi}\sin\theta~~,
 ~~f^{\prime}_{8\pi}= (A_{1\pi}\cos\theta +B_{2\pi})\sin\theta\nonumber \\
 f^{\prime}_{9\pi}& = & -\frac{E_{2}-M}{E_{\omega}}~~B_{1\pi}\sin^{2}
 \theta~~,~~
 f^{\prime}_{10\pi}=(A_{2\pi}\sin^{2}\theta + B_{3\pi})
 \nonumber \\
 f^{\prime}_{11\pi}& = & (A_{3\pi}\sin^{2}\theta + B_{4\pi})~~,~~
 f^{\prime}_{12\pi}=-(A_{1\pi}\cos\theta + B_{5\pi})\sin\theta~~,
 \nonumber \\
\end{eqnarray}
and
\begin{eqnarray}
 f_{i\sigma} & = & e~~\frac{g_{\omega\sigma\gamma}}{m_{\omega}}
 ~~\frac{g_{\sigma NN}}{t-m_{\sigma}^{2}}~~\sqrt{(E_{1}+M)~(E_{2}+M)}
 ~~f^{\prime}_{i\sigma}~~,\nonumber \\
 f^{\prime}_{1\sigma}& = & -(A_{3\sigma}\sin^{2}\theta + B_{2\sigma})~~,~~
 f^{\prime}_{2\sigma}=-(A_{4\sigma}\sin^{2}\theta + B_{3\sigma})\sin\theta
 \nonumber \\
 f^{\prime}_{3\sigma}& = & -(\frac{E_{\omega}}{|\vec{q}|}A_{4\sigma}
 \sin^{2}\theta + B_{4\sigma})~~,~~
 f^{\prime}_{4\sigma}=-(A_{2\sigma}\cos\theta +
 B_{6\sigma})\sin\theta\nonumber \\
 f^{\prime}_{5\sigma}& = & -A_{5\sigma}\sin^{2}\theta~~,~~
 f^{\prime}_{6\sigma}=-(-\frac{E_{\omega}}
 {|\vec{q}|}A_{4\sigma}\cos\theta + B_{5\sigma})\sin\theta\nonumber \\
 f^{\prime}_{7\sigma}& = &~~f^{\prime}_{8\sigma}~~=~~f^{\prime}
 _{9\sigma}~~=~~f^{\prime}_{10\sigma}~~=~~f^{\prime}_{11\sigma}~~
 =~~f^{\prime}_{12\sigma}~~=~~0~~,\nonumber \\
\end{eqnarray}
where $E_{1}(E_{2})$ is the energy of the initial(final) nucleon,
$E_{\omega}$ is the energy of $\omega$-meson, $\theta$ is the angle
between $\vec{k}$  and $\vec{q}$ in CMS, and the coefficients $A_{i\pi}$
$(i=1-3)$, $B_{i\pi}$ $(i=1-5)$, and $A_{i\sigma}$ $(i=3-5)$,
$B_{i\sigma}$ $(i=2-6)$ in Eqs. (7) and (8) are given in
Appendix A.

Similar expressions can also be written for the (s+u)-contributions to
the transversal amplitudes as

\begin{eqnarray}
 f_{is} & = & \frac{e}{W+M}~~\sqrt{(E_1+M)~(E_{2}+M)}
 ~~f^{\prime}_{is}~~,\nonumber \\
 f^{\prime}_{1s}& = & (A_{1s}+B_{1s}\cos\theta+C_{1s}\cos^{2}\theta)~~,~~
 f^{\prime}_{2s}= -(B_{1s}+C_{1s}\cos\theta)\sin\theta\nonumber \\
 f^{\prime}_{3s}& = & (A_{2s}+B_{1s}\cos\theta)~~,~~
 f^{\prime}_{4s}= (B_{2s}+C_{1s}\cos\theta)\sin\theta\nonumber \\
 f^{\prime}_{5s}& = & (-A_{2s}+B_{2s}\cos\theta+C_{1s}\cos^{2}\theta)~~,~~
 f^{\prime}_{6s}= B_{1s}\sin\theta\nonumber \\
 f^{\prime}_{7s}& = &~~f^{\prime}_{6s}~~,~~f^{\prime}_{8s}~~=~~-f^{\prime}_{4s}
 ~~,~~f^{\prime}_{9s}~~=~~-f^{\prime}_{5s}\nonumber \\
 f^{\prime}_{10s}& = &~~f^{\prime}_{3s}~~,~~f^{\prime}_{11s}~~=~~-f^{\prime}
 _{1s}~~,~~f^{\prime}_{12s}~~=~~-f^{\prime}_{2s}~~,\nonumber \\
\end{eqnarray}
\begin{eqnarray}
 f_{iu} & = & \frac{e}{u-M^{2}}~~\sqrt{(E_{1}+M)~(E_{2}+M)}
 ~~f^{\prime}_{iu}~~,\nonumber
\\
 f^{\prime}_{1u} & = & - (A_{1u}\sin^{2}\theta+ B_{1u})~~,~~
 f^{\prime}_{2u}=- (A_{1u}\cos\theta + B_{2u})~~\sin\theta\nonumber \\
 f^{\prime}_{3u} & = & -(A_{2u}\sin^{2}\theta+ B_{1u})~~,~~
 f^{\prime}_{4u}= (A_{3u}\cos\theta + B_{3u})~~\sin\theta\nonumber \\
 f^{\prime}_{5u} & = & -(A_{3u}\cos^{2}\theta+ B_{4u})~~,~~
 f^{\prime}_{6u}= A_{4u}~~\sin\theta\nonumber \\
 f^{\prime}_{7u} & = & -A_{5u}~~\sin\theta~~,~~
 f^{\prime}_{8u}= (A_{6u}\cos\theta+ B_{5u})~~\sin\theta\nonumber \\
 f^{\prime}_{9u} & = & (A_{7u}\sin^{2}\theta + B_{6u})~~,~~
 f^{\prime}_{10u}= (A_{8u}\sin^{2}\theta + B_{1u})\nonumber \\
 f^{\prime}_{11u} & = & -(A_{9u}\sin^{2}\theta+ B_{1u})~~,~~
 f^{\prime}_{12u}= -(A_{10u}\sin^{2}\theta + B_{7u})~~,\nonumber \\
\end{eqnarray}
where $W=\sqrt{s}$, the coefficients $A_{is}$ $(i=1,2)$,
$B_{is}$ $(i=1,2)$, $C_{is}$ $(i=1)$ and $A_{iu}$ $(i=1-10)$,
$B_{iu}$ $(i=1-7)$ in Eqs. (9) and (10) are given in
Appendix B and C, respectively.

\section{Numerical Results And Discussion}

In the suggested model there are two different sets of parameters,
namely the coupling constants, and the cut-off
parameters $\Lambda_{i}$ which characterize the t-dependence of
the phenomenological form factors $F_{\omega\pi\gamma}(t)$ and
$F_{\pi NN}(t)$ for the two vertexes of one-pion diagram:

\begin{eqnarray}
 F_{\omega\pi\gamma}(t)=\frac{\Lambda_{\omega\pi\gamma}^2-m_{\pi}^2}
 {\Lambda_{\omega\pi\gamma}^2-t}~~,~~
 F_{\pi NN}(t)=\frac{\Lambda_{\pi NN}^2-m_{\pi}^2}{\Lambda_{\pi NN
 }^2-t}~~,
\end{eqnarray}

Evidently, these two sets have different physical content and
different physical meaning. First of all, the parameters
$\Lambda_{i}$ are positive, whereas for the coupling constants
$g_{\omega NN}^{V}$ and $g_{\omega NN}^{T}$ not only their absolute values
are important but their signs as well, because of the essential
interference effects.
So, on the level of differential cross section with unpolarized
particles there is  a strong $\pi\bigotimes N$-interference, and
the interference of type $g_{\omega NN}^{V}g_{\omega NN}^{T}$, as
well. As a result, the
fitting procedure can produce not only the absolute values of both
constants $g_{\omega NN}^{V}$ and $g_{\omega NN}^{T}$ but their signs
also. Of course, it
is not the absolute signs we can speak here, but only about the relative
signs of coupling constants $g_{\omega NN}^{V}$ and $g_{\omega NN}^{}$
with respect to
the $\pi$-contribution. Therefore, it is possible to assume that
the product $g_{\omega\pi\gamma}g_{\pi NN}$ must be positive, thus
fixing by this agreement some system for relative signs. In any case, the
cut-off parameters $\Lambda_{i}$ must be positive
and can be fixed  at some plausible values.

So, in our model we have two fitting parameters, namely the
$\omega NN$-coupling constants $g_{\omega NN}^{V}$ and $g_{\omega
NN}^{T}$. To find these
constants we use the "new" experimental data about
$d\sigma(\gamma p\rightarrow p \omega)/dt$ in the near threshold
region \cite{R19}: namely, for $E_{\gamma}=$ $1.23$, $1.45$, $1.68$,
and $1.92$~GeV corresponding to four energy intervals, $1.1 <
E_{\gamma} < 1.35$~GeV,  $1.35 < E_{\gamma} < 1.55 $~GeV,
$1.55 < E_{\gamma} < 1.8$~GeV, and $1.8 < E_{\gamma} < 2.03$~GeV, and in
our fit we use all the experimental data in these energy intervals.
Minimizing procedure demonstrates that (s+u)-contribution,
being very important, can not be fixed uniquely on the basis of
existing experimental data about $d\sigma(\gamma p\rightarrow p
\omega)/dt$.  There are two sets of different pair $g_{\omega NN}^{V}$ and
$g_{\omega NN}^{T}$, which are equivalently good for the description of
differential cross section $d\sigma(\gamma p\rightarrow p
\omega)/dt$ with almost the same value of $\chi^{2}$. For
example, if one uses the "standard" values for the cut-off parameters,
namely $\Lambda_{\pi NN}=0.7$~GeV and
$\Lambda_{\omega\pi\gamma}=0.77$~GeV,
the best solution with $\chi^{2}/ndf=2.2$ corresponds to the following
values of $g_{\omega NN}^{V}$ and $g_{\omega NN}^{T}$:
\begin{eqnarray}
 (a)~~~g_{\omega NN}^{V}=-1.4~,~~g_{\omega NN}^{T}=0.4
\end{eqnarray}
To analyze the sensitivity of the "best" fit to  $\Lambda_{\pi
NN}$, and $\Lambda_{\omega\pi\gamma}$ we produce fitting with variable
values of $\Lambda_{i}$, and discover that the "standard" values of
$\Lambda_{i}$ are not the best ones. For example for $\Lambda_{\pi
NN}=0.5$~GeV, $\Lambda_{\omega\pi\gamma}=1.0$~GeV we find a better
solution,
namely
\begin{eqnarray}
 (b)~~~g_{\omega NN}^{V}=0.5~,~~g_{\omega NN}^{T}=0.1
\end{eqnarray}
with $\chi^{2}/ndf=1.6$. For these values of parameter $\Lambda_{i}$
the solution with negative value of coupling constant $g_{\omega NN}^{V}$,
namely
\begin{eqnarray}
 (c)~~~g_{\omega NN}^{V}=-0.4~,~~g_{\omega NN}^{T}=1.0
\end{eqnarray}
can also be found, but not with the best value of $\chi^{2}/ndf=2.5$,
which
is near to the solution (a). The resulting differential  cross sections
obtained using the above solutions of the coupling constants in the model
considered for $\gamma+p\rightarrow p+\omega$ at $E_{\gamma}=1.23$,
$1.45$, $1.68$, and $1.92$ GeV are shown in Fig. 2. All these
solutions are good enough to reproduce the t-dependence of $d\sigma/dt$
but they are different in physical content: the fit (b) is producing
positive $\pi\bigotimes N$-interference contribution to
$d\sigma/dt$ whereas the fit (a) and (c) negative
interference contribution. But in
all cases we are evidently improving in description of
t-behaviour for $-t>0.5$ $GeV^{2}$, in comparison with one-pion
exchange only. As we can see from Fig. 3,  the different sets  result
in different cross section for $\gamma n\rightarrow n \omega$. So
from the point of view of suggested model the future data about
$\gamma n\rightarrow n \omega$ will be very interesting.

In this respect the beam asymmetry $\Sigma$ which is very
sensitive to the considered variants of the model here, is also important.
Predicted behaviours of beam asymmetry $\Sigma$ for $\gamma+p\rightarrow
p+\omega$ and $\gamma+n\rightarrow  n+\omega$ are presented in Fig. 4 and
Fig. 5, respectively. In the case of (b) model, the one-nucleon
contribution is
producing in absolute value a large, in sign a negative value of
$\Sigma$, but the $\pi\bigotimes N$-interference is cancelling
this value. Therefore, we have here some "imitation" of pure
one-pion exchange, for which $\Sigma=0$ exactly, but for this set
of values of
corresponding coupling constants and cut-off parameters in this model
$\Sigma\neq 0$, being $\Sigma\leq 0.1$. In some sense contrary situation
appears for model (a) and (c), where the
one-nucleon contribution generates small values of $\Sigma$,
but the $\pi\bigotimes N$-interference is very important,
especially for the neutron target, producing even the maximal
value $|\Sigma|=1$ at $t\simeq 1.0 GeV^{2}$.

Another prediction of our model is the ratio of differential cross
sections $R=d\sigma(\gamma n\rightarrow n\omega)/d\sigma(\gamma
p\rightarrow p\omega)$ which is shown in Fig. 6. In this figure, different
contributions of exchange mechanisms to $d\sigma(\gamma
p\rightarrow p\omega)/dt$, R, $\Sigma(\gamma p\rightarrow p\omega)$ and
$\Sigma(\gamma n\rightarrow n\omega)$ at $E_{\gamma}=1.45$ GeV are shown
for the values of the $\omega NN$-coupling constants $g_{\omega
NN}^{V}=0.5$ and $g_{\omega NN}^{T}=0.1$.

In principle $\sigma$-contribution can be estimated here on
the basis of coupling constant $g_{\omega\sigma\gamma}$ obtained from the
branching ratio  $\omega\rightarrow\pi^{0}+\pi^{0}+\gamma$.
Considering two mechanisms for this decay namely,
$\sigma$-exchange:
$\omega\rightarrow\sigma+\gamma\rightarrow\pi^{0}+\pi^{0}+\gamma$,
and $\rho$-exchange:
$\omega\rightarrow\pi^{0}+\rho^{0}\rightarrow\pi^{0}+\pi^{0}+\gamma$, then
utilizing the experimental value of branching ratio
it is possible to find two solutions for
$g_{\omega\sigma\gamma}$\cite{R20}. But these possible solutions for
$g_{\omega\sigma\gamma}$ with different signs will result
however, in very small contribution to $d\sigma/dt$ and $\Sigma$,
on proton and neutron targets.

Of course, our investigation is succeptible to both experimental
and theoretical uncertainties. Experimentally, a systematic study
of differential cross section $d\sigma(\gamma p\rightarrow p
\omega)/dt$ with high enough accuracy is not available. And
the absence of any polarization data about process $\gamma
+p\rightarrow p+\omega$ seems as a serious defect at the moment.
This, combined with overall poor quality of the reported data, may
make a detailed analysis non-conclusive at this stage. Therefore,
our calculations are performed on the boundary of the modern approaches
to these processes, and as such should be considered as a
first approach.

Although our $d\sigma/dt$ fit demonstrates our point that the
existing data about real $\omega$-photoproduction in the near
threshold region can be explained in the framework of
($\pi+N$)-model, we do not consider our success to be decisive.
Indeed, we obtained a fit in which only nucleon exchange in s- and
u-channels is taken into account. In principle a fit of better
quality can be done in a model with $N^{*}$-contributions. But we
must repeat once more that the quality of existing data is not so
good for more refined analysis. In any case it is demonstrated here
that the proposed model in this work provides not only explanation
of
existing data about $d\sigma(\gamma p\rightarrow p \omega)/dt$ in
the near threshold region in the whole t region, but our analysis
also proves that information about polarization
observables in $\gamma p\rightarrow p \omega$ will help in
clarifying the picture of $\omega$-meson photoproduction mechanism.

\section{Conclusions}

So our previous analysis allows us to obtain the following
conclusions:

$\bullet$ The existing experimental data about t-dependence of the
differential cross section $d\sigma(\gamma p\rightarrow p
\omega)/dt$ in the near threshold region ($E_{\gamma}\leq 2.0$~GeV)
can be described in the framework of model with $\pi$- and
N-exchanges, only.

$\bullet$ For the coupling constants $g_{\omega NN}^{V}$ and $g_{\omega
NN}^{T}$ of the
$\omega NN$-vertex the different solutions have been obtained,
corresponding to positive and negative values of $g_{\omega NN}^{V}$ and
$g_{\omega NN}^{T}$,
respectively, with constructive and destructive $\pi\bigotimes
N$-interference contributions to the differential cross section
$d\sigma(\gamma p\rightarrow p \omega)/dt$. Let us note that all these
sets of
the coupling constants $g_{\omega NN}^{V}$ and $g_{\omega NN}^{T}$ are
different from the
"standard" values of these constants for the space-like values of
vector meson four-momentum.

$\bullet$ It is demonstrated that the t-behaviour of the beam
asymmetry $\Sigma$ is especially sensitive to above mentioned sets
of $\omega NN$-coupling constants obtained in time-like region of vector
meson four momentum.

\section*{Acknowledgement}

We thank M. P. Rekalo for suggesting this problem to us and we gratefully
acknowledge his guidance and fruitful discussions during the course
of our work.

\newpage
\appendix

\section{The coefficients in transversal amplitudes of $\pi$- and $\sigma$
-exchange for t-channel}
\begin{eqnarray}
 A_{1\pi} & = & -\frac{(t-2M^{2}+2E_{1}E_{2})(E_{2}-M)}{2 E_{\omega}}
 \nonumber \\
 A_{2\pi} & = & |\vec{k}|~~(E_{2}-M)\nonumber \\
 A_{3\pi} & = & -\left(A_{1\pi}+A_{2\pi}+\frac{|\vec{q}|^{2}(E_{1}-M)}
 {E_{\omega}}\right)\nonumber \\
 A_{4\pi} & = & A_{2\pi}+E_{\omega}(E_{1}-M)\nonumber \\
 A_{5\pi} & = & \frac{(t-2M^{2}+2E_{1}E_{2})}{2}\left(\frac{|\vec{k}|}
 {(E_{1}+M)}+\frac{E_{\omega}}{(E_{2}+M)}\right)\nonumber \\
 B_{1\pi} & = & |\vec{k}|E_{\omega}-\frac{(t-2M^{2}+2E_{1}E_{2})}{2}\nonumber \\
 B_{2\pi} & = & \frac{|\vec{q}|}{(E_{2}+M)}\left((t-2M^{2}+2E_{1}E_{2})-
 \frac{|\vec{k}| m_{\omega}^{2}}{E_{\omega}}\right)\nonumber \\
 B_{3\pi} & = &  A_{5\pi}- A_{4\pi}\nonumber \\
 B_{4\pi} & = &
 A_{4\pi}-\frac{(t-2M^{2}+2E_{1}E_{2})}{2}\left(\frac{|\vec{k}|}
 {(E_{1}+M)}+\frac{(E_{2}-M)}{E_{\omega}}+\frac{m_{\omega}^{2}}{E_{\omega}
 (E_{2}+M)}\right)\nonumber \\
 B_{5\pi} & = & |\vec{q}|\left(\frac{A_{5\pi}}{E_{\omega}}-(E_{1}-M)\right)
 \nonumber \\
 A_{1\sigma} & = & \frac{|\vec{k}||\vec{q}|^{2}}{E_{\omega}}
 +(E_{1}-M)(E_{2}-M)\nonumber \\
 A_{2\sigma} & = & \frac{(t-2M^{2}+2E_{1}E_{2})(E_{2}-M)|\vec{k}|}
 {2 E_{\omega}(E_{1}+M)}\nonumber \\
 A_{3\sigma} & = & A_{2\sigma}-A_{1\sigma}\nonumber \\
 A_{4\sigma} & = & \frac{(E_{1}-M)(E_{2}-M)|\vec{q}|}{E_{\omega}}
 \nonumber \\
 A_{5\sigma}& = & \frac{(m_{\omega}^{2}-t)(E_{1}-M)(E_{2}-M)}{2
 E_{\omega}|\vec{k}|}\nonumber \\
 B_{1\sigma} & = & \left(\frac{(t-2M^{2}+2E_{1}E_{2})}{2}\right)
 \left(1+\frac{|\vec{k}|E_{\omega}}{(E_{1}+M)(E_{2}+M)}\right)
 \nonumber \\
 B_{2\sigma} & = & A_{1\sigma}-B_{1\sigma}+ \frac{|\vec{k}|m_{\omega}^{2}}
 {E_{\omega}}\nonumber \\
 B_{3\sigma} & = & \frac{|\vec{q}|}{E_{\omega}}(B_{1\sigma}-
 |\vec{k}|E_{\omega})-A_{4\sigma}\nonumber \\
 B_{4\sigma} & = &  B_{1\sigma}-|\vec{k}|E_{\omega}-(E_{1}-M)(E_{2}-M)
 \nonumber \\
 B_{5\sigma} & = & \frac{E{\omega}(E_{1}-M)|\vec{q}|}{(E_{2}+M)}
 \nonumber \\
 B_{6\sigma} & = & \frac{B_{5\sigma}m_{\omega}^{2}}{E_{\omega}^{2}}-
 \frac{(t-2M^{2}+2E_{1}E_{2})|\vec{k}||\vec{q}|}{2(E_{1}+M)(E_{2}+M)}
\end{eqnarray}
\section{The coefficients in transversal amplitudes for s-channel}
\begin{eqnarray}
 A_{1s} & = & -\frac{1}{E_{\omega}}\left(Q_{N}-\frac{\kappa_{N}}
 {2M}(W-M)\right)\left(g_{\omega NN}^{V}(W-M)-\frac{g_{\omega NN}^{T}}{2M}
 m_{\omega}^{2}\right)\nonumber \\
 A_{2s} & = & -\left(Q_{N}-\frac{\kappa_{N}}{2M}(W-M)\right)\left(g_{\omega NN}
 ^{V}-\frac{g_{\omega NN}^{T}}{2M}(W-M)\right)\nonumber \\
 B_{1s} & = & -\left(Q_{N}+\frac{\kappa_{N}}{2M}(W+M)\right)\left(g_{\omega NN}
 ^{V}+\frac{g_{\omega NN}^{T}}{2M}(W+M)\right)\nonumber \\
 & \times & \left(\frac{|\vec{k}||\vec{q}|(W+M)}{(E_{1}+M)(E_{2}+M)
 (W-M)}\right)\nonumber \\
 B_{2s} & = & -\left(Q_{N}+\frac{\kappa_{N}}{2M}(W+M)\right)
 \left(g_{\omega NN}^{V}(W+M)+\frac{g_{\omega NN}^{T}}
 {2M}m_{\omega}^{2}\right)\nonumber \\
 & \times & -\left(\frac{|\vec{k}||\vec{q}|(W+M)}{E_{\omega}(E_{1}+M)
 (E_{2}+M)(W-M)}\right)\nonumber \\
 C_{1s} & = & \frac{(E_{2}-M)}{E_{\omega}}
 \left(Q_{N}-\frac{\kappa_{N}}{2M}(W-M)\right)\left(g_{\omega NN}
 ^{V}+\frac{g_{\omega NN}^{T}}{2M}(W+M)\right)\nonumber \\
\end{eqnarray}
\section{The coefficients in transversal amplitudes for u-channel}
\begin{eqnarray}
 A_{1u} & = & \left[Q_{N}(g_{\omega NN}^{V}a_{1u}+
 \frac{g_{\omega NN}^{T}}{2M}a_{3u})-
 \frac{\kappa_{N}}{2M}a_{5u}(g_{\omega NN}^{V}
 +\frac{g_{\omega NN}^{T}}{2M}(W+M))\right]\nonumber \\
 A_{2u} & = & \left[Q_{N}\frac{g_{\omega NN}^{T}}{2M}+\frac{\kappa_{N}}
 {2M}(g_{\omega NN}^{V}-\frac{g_{\omega NN}^{T}}{2M}(W-M))
 \right]a_{11u}\nonumber \\
 A_{3u} & = & \left[Q_{N}(g_{\omega NN}^{V}a_{12u}+
 \frac{g_{\omega NN}^{T}}{2M}a_{13u})-
 \frac{\kappa_{N}}{2M}(g_{\omega NN}^{V}a_{14u}+
 \frac{g_{\omega NN}^{T}}{2M}a_{15u})
 \right]\nonumber \\
 A_{4u} & = & \left[Q_{N}(g_{\omega NN}^{V}a_{20u}
 +\frac{g_{\omega NN}^{T}}{2M}a_{21u})+
 \frac{\kappa_{N}}{2M}(g_{\omega NN}^{V}a_{22u}
 -\frac{g_{\omega NN}^{T}}{2M}~a_{23u})\right]\nonumber \\
 A_{5u} & = & \left[Q_{N}(g_{\omega NN}^{V}a_{24u}
 +\frac{g_{\omega NN}^{T}}{2M}a_{25u})+
 \frac{\kappa_{N}}{2M}a_{26u}(g_{\omega NN}^{V}+
 \frac{g_{\omega NN}^{T}}{2M}(W+M))\right]\nonumber \\
 A_{6u} & = & \left[Q_{N}\frac{g_{\omega NN}^{T}}{2M}a_{28u}+
 \frac{\kappa_{N}}{2M}(g_{\omega NN}^{V}a_{30u}-\frac{g_{\omega NN}^{T}}
 {2M}a_{32u})\right]\nonumber \\
 A_{7u} & = & \left[-Q_{N}(g_{\omega NN}^{V}a_{35u}
 +\frac{g_{\omega NN}^{T}}{2M}a_{37u})+
 \frac{\kappa_{N}}{2M}(g_{\omega NN}^{V}a_{39u}+
 \frac{g_{\omega NN}^{T}}{2M}a_{40u})\right]\nonumber \\
 A_{8u} & = & \left[Q_{N}(g_{\omega NN}^{V}
 +\frac{g_{\omega NN}^{T}}{2M}(E_{1}+M))-
 \frac{\kappa_{N}}{2M}|\vec{k}|(g_{\omega NN}^{V}+
 \frac{g_{\omega NN}^{T}}{2M}(W+M))\right]a_{41u}\nonumber \\
 A_{9u} & = & \left[Q_{N}(g_{\omega NN}^{V}a_{35u}
 -\frac{g_{\omega NN}^{T}}{2M}a_{42u})+
 \frac{\kappa_{N}}{2M}(g_{\omega NN}^{V}a_{43u}-
 \frac{g_{\omega NN}^{T}}{2M}a_{44u})\right]\nonumber \\
 A_{10u} & = & \left[Q_{N}\frac{g_{\omega NN}^{T}}{2M}a_{42u}+
 \frac{\kappa_{N}}{2M}(g_{\omega NN}^{V}a_{43u}-
 \frac{g_{\omega NN}^{T}}{2M}a_{44u})\right]\nonumber \\
 B_{1u} & = & \left[Q_{N}(-g_{\omega NN}^{V}a_{2u}
 +\frac{g_{\omega NN}^{T}}{2M}a_{4u})-
 \frac{\kappa_{N}}{2M}(g_{\omega NN}^{V}a_{6u}-
 \frac{g_{\omega NN}^{T}}{2M}a_{7u})\right]\nonumber \\
 B_{2u} & = & \left[Q_{N}(g_{\omega NN}^{V}a_{8u}
 -\frac{g_{\omega NN}^{T}}{2M}a_{9u})-
 \frac{\kappa_{N}}{2M}a_{10u}(g_{\omega NN}^{V}+
 \frac{g_{\omega NN}^{T}}{2M}(W+M))\right]\nonumber \\
 B_{3u} & = & \frac{1}{E_{\omega}}\left[Q_{N}(-g_{\omega NN}^{V}a_{9u}
 +\frac{g_{\omega NN}^{T}}{2M}a_{8u}m_{\omega}^{2})-
 \frac{\kappa_{N}}{2M}a_{10u}(g_{\omega NN}^{V}(W+M)+
 \frac{g_{\omega NN}^{T}}{2M}m_{\omega}^{2})\right]\nonumber \\
 B_{4u} & = & \left[Q_{N}(-g_{\omega NN}^{V}a_{16u}
 +\frac{g_{\omega NN}^{T}}{2M}a_{17u})-
 \frac{\kappa_{N}}{2M}(g_{\omega NN}^{V}a_{18u}-
 \frac{g_{\omega NN}^{T}}{2M}a_{19u})\right]\nonumber \\
 B_{5u} & = & \left[Q_{N}(g_{\omega NN}^{V}a_{27u}
 +\frac{g_{\omega NN}^{T}}{2M}a_{29u})+
 \frac{\kappa_{N}}{2M}(g_{\omega NN}^{V}a_{31u}-
 \frac{g_{\omega NN}^{T}}{2M}a_{33u})\right]\nonumber \\
 B_{6u} & = & \left[Q_{N}(g_{\omega NN}^{V}a_{34u}
 +\frac{g_{\omega NN}^{T}}{2M}a_{36u})-
 \frac{\kappa_{N}}{2M}(g_{\omega NN}^{V}a_{38u}-
 \frac{g_{\omega NN}^{T}}{2M}a_{6u}\frac{m_{\omega}^{2}}{E_{\omega}})
 \right]\nonumber \\
 B_{7u} & = & \left[Q_{N}(g_{\omega NN}^{V}a_{45u}
 -\frac{g_{\omega NN}^{T}}{2M}a_{46u})+
 \frac{\kappa_{N}}{2M}(g_{\omega NN}^{V}a_{47u}-
 \frac{g_{\omega NN}^{T}}{2M}a_{48u})\right]\nonumber \\
\end{eqnarray}
where
\begin{eqnarray}
 a_{1u}& = & \frac{(E_{2}-M)}{E_{\omega}}
 \left[\frac{(t-2M^{2}+2E_{1}E_{2})}{(E_{1}+M)}+(2E_{1}+W+M)\right]
 \nonumber \\
 a_{2u}& = & \left[\frac{(t-2M^{2}+2E_{1}E_{2})(W+M)}
 {2(E_{1}+M)(E_{2}+M)}+(W-M)\right]\nonumber \\
 a_{3u}& = & \frac{(E_{2}-M)}{E_{\omega}}
 \left[(t+m_{\omega}^{2}+2|\vec{k}|E_{\omega})
 -(E_{1}-M-|\vec{k}|)(E_{2}+M-E_{\omega})\right]
 \nonumber \\
 a_{4u}& = & \left[\frac{(t-2M^{2}+2E_{1}E_{2})
 (E_{1}+M-|\vec{k}|)(E_{2}+M-E_{\omega})}{2(E_{1}+M)(E_{2}+M)}
 \right]\nonumber \\
 & - & (E_{1}-M-|\vec{k}|)(E_{2}-M-E_{\omega})\nonumber \\
 a_{5u}& = & \frac{(E_{2}-M)}{E_{\omega}}
 \left[\frac{(t-2M^{2}+2E_{1}E_{2})|\vec{k}|}
 {(E_{1}+M)}+2E_{2}(E_{1}-M)+2|\vec{k}|E_{\omega}
 \right]\nonumber \\
 a_{6u}& = & \left[\left(\frac{|\vec{k}|}{(E_{1}+M)}-
 \frac{E_{2}}{(E_{2}+M)}\right)(t-2M^{2}+2E_{1}E_{2})\right]
 \nonumber \\
 & - & 2|\vec{k}|(E_{2}-M)+2E_{2}(E_{1}-M)\nonumber \\
 a_{7u}& = & \left[\left(\frac{|\vec{k}|(W-M)}{(E_{1}+M)}+
 \frac{E_{2}(W+M)}{(E_{2}+M)}\right)(t-2M^{2}+2E_{1}E_{2})
 \right]\nonumber \\
 & + & 2|\vec{k}|(E_{2}-M)(W+M)+2E_{2}(E_{1}-M)(W-M)\nonumber \\
 a_{8u}& = & \left[\frac{|\vec{k}||\vec{q}|}{(E_{1}+M)}
 \left(2+\frac{(E_{1}+M-|\vec{k}|)}{(E_{2}+M)}\right)\right]
 \nonumber \\
 a_{9u}& = & \left[\frac{|\vec{k}||\vec{q}|(E_{2}+M-E_{\omega})}
 {(E_{2}+M)}+\frac{|2\vec{k}||\vec{q}|(W-M)}{(E_{1}+M)}
 -\frac{(E_{1}-M)(W+M)|\vec{q}|}{(E_{2}+M)}\right]\nonumber \\
 a_{10u}& = & \frac{2|\vec{k}||\vec{q}|E_{2}}{(E_{2}+M)}
 \nonumber \\
 a_{11u}& = & 2(E_{1}-M)(E_{2}-M)\nonumber \\
 a_{12u}& = & (E_{2}-M)\left[\left(\frac{(t-2M^{2}+2E_{1}E_{2})}
 {E_{\omega}(E_{1}+M)}\right)+\left(2-\frac{|\vec{k}|
 (E_{1}+M-|\vec{k}|)}{E_{\omega}(E_{1}+M)}\right)
 \right]\nonumber \\
 a_{13u}& = & \frac{(E_{2}-M)}{E_{\omega}}
 \left[(t-2M^{2}+2E_{1}E_{2})+(W+M)(2E_{\omega}-|\vec{k}|)
 -(E_{1}-M)(E_{2}+M-E_{\omega})\right]\nonumber \\
 a_{14u}& = & \left[\frac{(t-2M^{2}+2E_{1}E_{2})(E_{2}-M)|\vec{k}|}
 {E_{\omega}(E_{1}+M)}+
 \frac{2|\vec{k}|(W+M)(E_{2}-M)}{E_{\omega}}\right]\nonumber \\
 & + &
 \frac{2E_{2}(E_{1}-M)(E_{2}-M)}{E_{\omega}}\nonumber \\
 a_{15u}& = & \left[\frac{(t-2M^{2}+2E_{1}E_{2})(E_{2}-M)
 |\vec{k}|(W+m)}{E_{\omega}(E_{1}+M)}+
 \frac{2|\vec{k}|m_{\omega}^{2}(E_{2}-M)}{E_{\omega}}\right]
 \nonumber \\
 & + &
 \frac{2E_{2}(E_{1}-M)(E_{2}-M)(W+M)}{E_{\omega}}\nonumber \\
 a_{16u}& = & \left[\frac{(t-2M^{2}+2E_{1}E_{2})}
 {E_{\omega}(E_{1}+M)}\left(\frac{(E_{2}+M-
 E_{\omega})(E_{1}+M-|\vec{k}|)}{2(E_{2}+M)}+(E_{2}-M)\right)
 \right]\nonumber \\
 & + &
 2(E_{2}-M)+(E_{1}-M-|\vec{k}|)\nonumber \\
 a_{17u}& = & \left[\frac{(t-2M^{2}+2E_{1}E_{2})}
 {E_{\omega}}\left(\frac{(W+M)m_{\omega}^{2}}
 {2(E_{1}+M)(E_{2}+M)}-(E_{2}-M)\right)
 \right]\nonumber \\
 & - &
 2(E_{2}-M)(W+M)+E_{\omega}(W-M)-2(E_{2}-M)(E_{1}-M-|\vec{k}|)
 \nonumber \\
 a_{18u}& = & \left[\frac{(t-2M^{2}+2E_{1}E_{2})}
 {E_{\omega}}\left(\frac{|\vec{k}|E_{\omega}}
 {(E_{1}+M)}+\frac{E_{2}(W+M)}{(E_{2}+M)}\right)
 \right]\nonumber \\
 & + &
 2E_{2}(E_{1}-M)\nonumber \\
 a_{19u}& = & \left[(t-2M^{2}+2E_{1}E_{2})
 \left(\frac{|\vec{k}|(W-M)}
 {(E_{1}+M)}-\frac{E_{2}m_{\omega}^{2}}{E_{\omega}(E_{2}+M)}
 \right)\right]\nonumber \\
 & + &
 2E_{2}(E_{1}-M)(W-M)\nonumber \\
 a_{20u}& = & \frac{|\vec{k}||\vec{q}|(W+M)}{(E_{1}+M)(E_{2}+M)}
 \nonumber \\
 a_{21u}& = & \frac{|\vec{q}|}{(E_{2}+M)}
 \left[\frac{(t-2M^{2}+2E_{1}E_{2})|\vec{k}|}{(E_{1}+M)}
 +|\vec{k}|(W+M)\right]\nonumber \\
 & + &
 \frac{|\vec{q}|(E_{1}-M)(E_{2}+M-E_{\omega})}{(E_{2}+M)}
 \nonumber \\
 a_{22u}& = & \frac{|\vec{q}|}{(E_{2}+M)}
 \left[\frac{(t-2M^{2}+2E_{1}E_{2})|\vec{k}|}{(E_{1}+M)}
 +2M|\vec{k}|+2(E_{1}-M)(E_{2}+M)\right]\nonumber \\
 a_{23u}& = & \frac{|\vec{q}|}{(E_{2}+M)}
 \left[\frac{(t-2M^{2}+2E_{1}E_{2})|\vec{k}|(W-M)}{(E_{1}+M)}
 -2M|\vec{k}|(W+M)\right]\nonumber \\
 & + &
 2|\vec{q}|(E_{1}-M)(W-M)\nonumber \\
 a_{24u}& = & \frac{|\vec{k}||\vec{q}|}{(E_{1}+M)(E_{2}+M)}
 \left[\frac{(t-2M^{2}+2E_{1}E_{2})}{(E_{1}-M)}
 -2(E_{2}+M)+(W+M)\right]\nonumber \\
 a_{25u}& = & \frac{|\vec{k}||\vec{q}|}{(E_{1}+M)(E_{2}+M)}
 \left[\frac{(t-2M^{2}+2E_{1}E_{2})(E_{1}+M)}{(E_{1}-M)}
 +(W-3M)(E_{2}+M) \right]\nonumber \\
 & + &
 \frac{|\vec{k}||\vec{q}|E_{\omega}(E_{1}+M-|\vec{k}|)}
 {(E_{1}+M)(E_{2}+M)}\nonumber \\
 a_{26u}& = & \frac{|\vec{q}|}{(E_{2}+M)}
 \left[(t-2M^{2}+2E_{1}E_{2})+2|\vec{k}|E_{2}\right]\nonumber \\
 a_{27u}& = & \frac{|\vec{q}|}{E_{\omega}(E_{2}+M)}
 \left[\frac{(t-2M^{2}+2E_{1}E_{2})(E_{1}+M-|\vec{k}|)}
 {2(E_{1}+M)}+|\vec{k}|(W+M)\right]\nonumber \\
 & - &
 \frac{|\vec{q}|(E_{1}-M)(E_{2}+M-E_{\omega})}
 {E_{\omega}(E_{2}+M)}\nonumber \\
 a_{28u}& = & \frac{(E_{2}-M)}{E_{\omega}}
 \left[(t-2M^{2}+2E_{1}E_{2})+(E_{1}-M)(W+M)+
 |\vec{k}|(E_{2}+M-E_{\omega})\right]\nonumber \\
 a_{29u}& = & \frac{|\vec{k}||\vec{q}|m_{\omega}^{2}
 (E_{1}+M-|\vec{k}|)}
 {E_{\omega}(E_{1}+M)(E_{2}+M)}\nonumber \\
 a_{30u}& = & \frac{(E_{2}-M)}{E_{\omega}}
 \left[(t-2M^{2}+2E_{1}E_{2})+2|\vec{k}|(E_{2}+M)+
 2W(E_{1}-M)\right]\nonumber \\
 a_{31u}& = & \left[\frac{2|\vec{k}||\vec{q}|}{E_{\omega}}
 \left(\frac{|\vec{k}|(W-M)}{(E_{1}+M)}+\frac{M(W+M)}{(E_{2}+M)}
 \right)\right]\nonumber \\
 a_{32u}& = & \frac{(t-2M^{2}+2E_{1}E_{2})(E_{2}-M)(W-M)}
 {E_{\omega}}\nonumber \\
 a_{33u}& = & \frac{|\vec{q}|}{E_{\omega}}
 \left[\left((W-M)-\frac{W|\vec{k}|(E_{2}+M-E_{\omega})}
 {(E_{1}+M)(E_{2}+M)}\right)(t-2M^{2}+2E_{1}E_{2})\right]
 \nonumber\\
 & + &
 \frac{2|\vec{q}|m_{\omega}^{2}}{E_{\omega}}
 \left((E_{1}-M)-\frac{M|\vec{k}|}{(E_{2}+M)}\right)\nonumber \\
 a_{34u}& = & \left[\frac{(E_{1}+M-|\vec{k}|)}
 {E_{\omega}(E_{1}+M)}\left(\frac{(t-2M^{2}+2E_{1}E_{2})
 (E_{2}+M-E_{\omega})}{2(E_{2}+M)}+|\vec{k}|(E_{2}-M)\right)
 \right]\nonumber \\
 & + & (E_{1}-M-|\vec{k}|)\nonumber \\
 a_{35u}& = & \frac{|\vec{k}|(E_{2}-M)(E_{1}+M-|\vec{k}|)}
 {E_{\omega}(E_{1}+M)}\nonumber \\
 a_{36u}& = & \left[\frac{|\vec{k}|(E_{2}-M)(E_{1}+M-|\vec{k}|)}
 {(E_{1}+M)}+\frac{(E_{1}-M)(E_{2}-M)(W+M)}{E_{\omega}}\right]
 \nonumber \\
 & + &
 \left[\frac{|\vec{k}|(E_{2}-M)(E_{2}+M-E_{\omega})}{E_{\omega}}
 -\frac{(t-2M^{2}+2E_{1}E_{2})m_{\omega}^{2}(W+M)}
 {2E_{\omega}(E_{1}+M)(E_{2}+M)}\right]\nonumber \\
 & - & E_{\omega}(W-M)\nonumber \\
 a_{37u}& = & \frac{(E_{2}-M)}{E_{\omega}}
 \left[(t-2M^{2}+2E_{1}E_{2})+(E_{1}-M)(W+M)+
 |\vec{k}|(E_{2}+M-E_{\omega})\right]\nonumber \\
 a_{38u}& = & \frac{(E_{2}-M)}{E_{\omega}}
 \left[(t-2M^{2}+2E_{1}E_{2})+2(E_{1}-M)(E_{\omega}+M)
 \right]\nonumber \\
 a_{39u}& = & \left[\frac{(t-2M^{2}+2E_{1}E_{2})}
 {E_{\omega}}\left(\frac{|\vec{k}|(W-M)}{(E_{1}+M)}
 +\frac{E_{2}(W+M)}{(E_{2}+M)}\right)\right]\nonumber \\
 & + &
 2|\vec{k}|(E_{2}-M)+2E_{2}(E_{1}-M)+\frac{2M(E_{1}-M)(E_{2}-M)}
 {E_{\omega}}\nonumber \\
 a_{40u}& = & \frac{(E_{2}-M)}{E_{\omega}}
 \left[\frac{(t-2M^{2}+2E_{1}E_{2})(W-M)}{2}+2|\vec{k}|(E_{2}+M)
 (W-M)\right]\nonumber \\
 & - &
 \frac{2W(E_{2}-M)(E_{1}-M)(E_{2}+M-E_{\omega})}{E_{\omega}}
 \nonumber \\
 a_{41u}& = & 2(E_{2}-M)\nonumber \\
 a_{42u}& = & \left[\frac{(E_{2}-M)(t-2M^{2}+2E_{1}E_{2})
 (E_{1}+M)}{E_{\omega}(E_{1}+M)}\right]\nonumber \\
 & + &
 \left[\frac{(E_{2}-M)|\vec{k}|(E_{2}+M-E_{\omega})
 (E_{1}+M-|\vec{k}|)}{E_{\omega}(E_{1}+M)}\right]\nonumber \\
 a_{43u}& = & \frac{(E_{2}-M)}{E_{\omega}}
 \left[(t-2M^{2}+2E_{1}E_{2})+
 2|\vec{k}|(E_{2}+M)\right]\nonumber \\
 & + &
 \frac{2(E_{2}-M)(E_{1}-M)(E_{\omega}-M)}{E_{\omega}}\nonumber \\
 a_{44u}& = & \frac{(E_{2}-M)(W-m)}{E_{\omega}}
 \left[(t-2M^{2}+2E_{1}E_{2})+2|\vec{k}|(E_{2}+M)\right]
 \nonumber \\
 & + &
 \frac{2(E_{2}-M)(E_{1}-M)(M(E_{2}+M-E_{\omega})+E_{\omega}(W+M))}
 {E_{\omega}}\nonumber \\
 a_{45u}& = & \frac{|\vec{q}|(E_{1}+M-|\vec{k}|)}
 {E_{\omega}(E_{1}+M)(E_{2}+M)}\left[\frac{(t-2M^{2}+2E_{1}E_{2})}
 {2}-|\vec{k}|E_{\omega}\right]\nonumber \\
 a_{46u}& = & \frac{|\vec{q}|}{(E_{2}+M)}
 \left[|\vec{k}|(W+M)-(E_{1}-M)(E_{2}+M-E_{\omega})\right]
 \nonumber \\
 a_{47u}& = & 2|\vec{q}|\left[(E_{1}-M)-\frac{2M|\vec{k}|}
 {(E_{2}+M)}\right]\nonumber \\
 a_{48u}& = & 2|\vec{k}||\vec{q}|\left[\frac{|\vec{k}|(W-M)}
 {(E_{1}+M)}+\frac{M(W+M)}{(E_{2}+M)}\right]\nonumber \\
\end{eqnarray}
\newpage
\begin{figure}
$\left. \right.$
\vskip 15cm
    \includegraphics{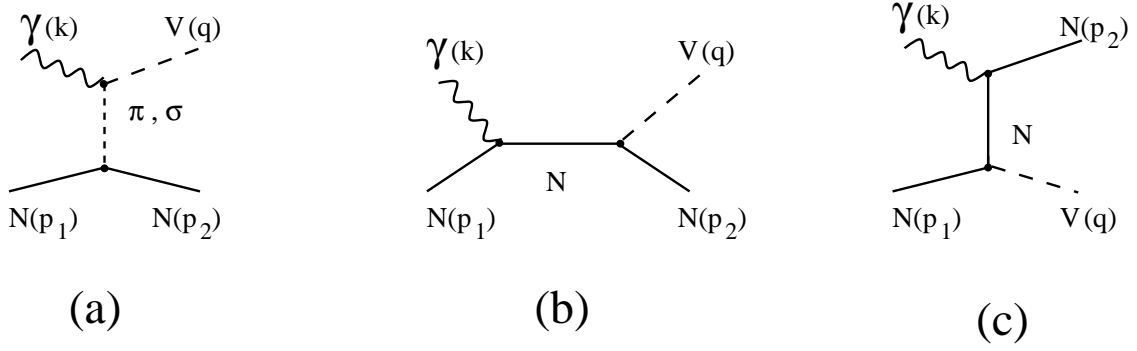}
\vskip -1cm
\caption{Mechanisms of the model for
$\omega$-photoproduction: (a) t-channel exchanges, (b) and (c)
s- and u-channel nucleon exchanges. }
\end{figure}
\newpage
\begin{figure}
$\left. \right.$
\vskip 10cm
    \includegraphics{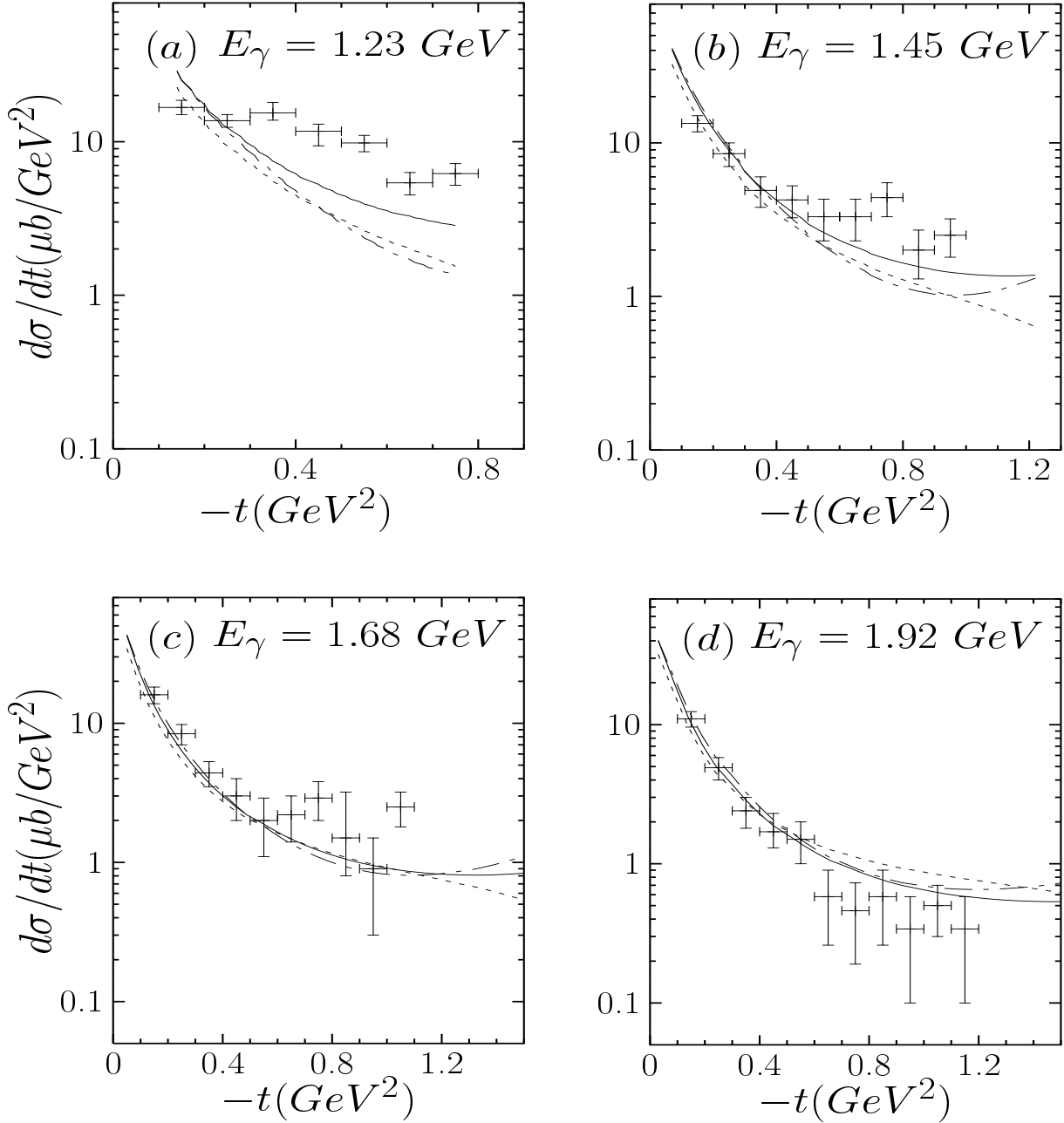}
\vskip 8.5 cm
\caption{Comparison of experimental differential cross section data for
$\gamma+p \rightarrow p+\omega$ at $E_{\gamma}=$ $1.23$,
$1.45$, $1.68$ and $1.92$~GeV from [19]
with the calculation of suggested model. Solid, dashed and
dot-dashed lines correspond to $g_{\omega NN}^{V}=0.5$,
~$g_{\omega NN}^{T}=0.1$~; $g_{\omega NN}^{V}=-0.4$,
~$g_{\omega NN}^{T}=1.0$~; and $g_{\omega NN}^{V}=-1.4$,
~$g_{\omega NN}^{T}=0.4$, respectively.}
\end{figure}
\newpage
\begin{figure}
$\left. \right.$
\vskip 10cm
    \includegraphics{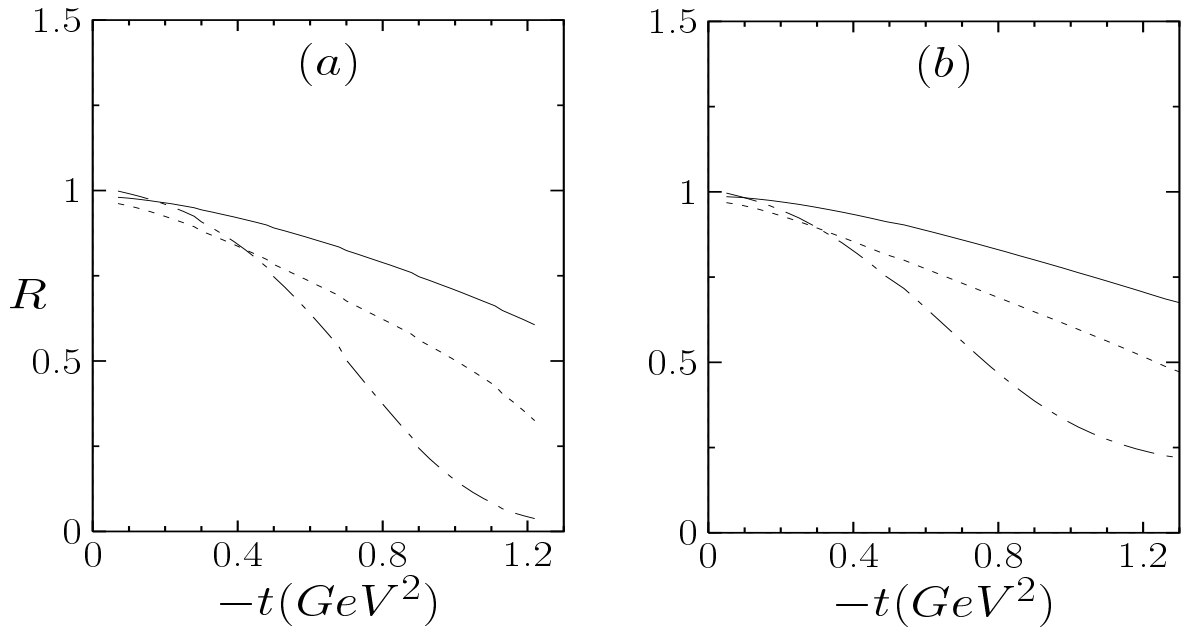}
\vskip 5.5cm
\caption{Ratio of differential cross section on neutron and
proton target ($R=$ $d\sigma(\gamma n\rightarrow n
\omega/$ $d\sigma(\gamma p\rightarrow p \omega)$ ) at
(a) $E_{\gamma}=$ $1.45$~GeV and (b) $E_{\gamma}=$
$1.68$~GeV with the total contributions
of exchange mechanisms ($\pi$,s, u). Notation for different graphs is
the same as in Fig.2.}
\end{figure}
\newpage
\begin{figure}
$\left. \right.$
\vskip 10cm
    \includegraphics{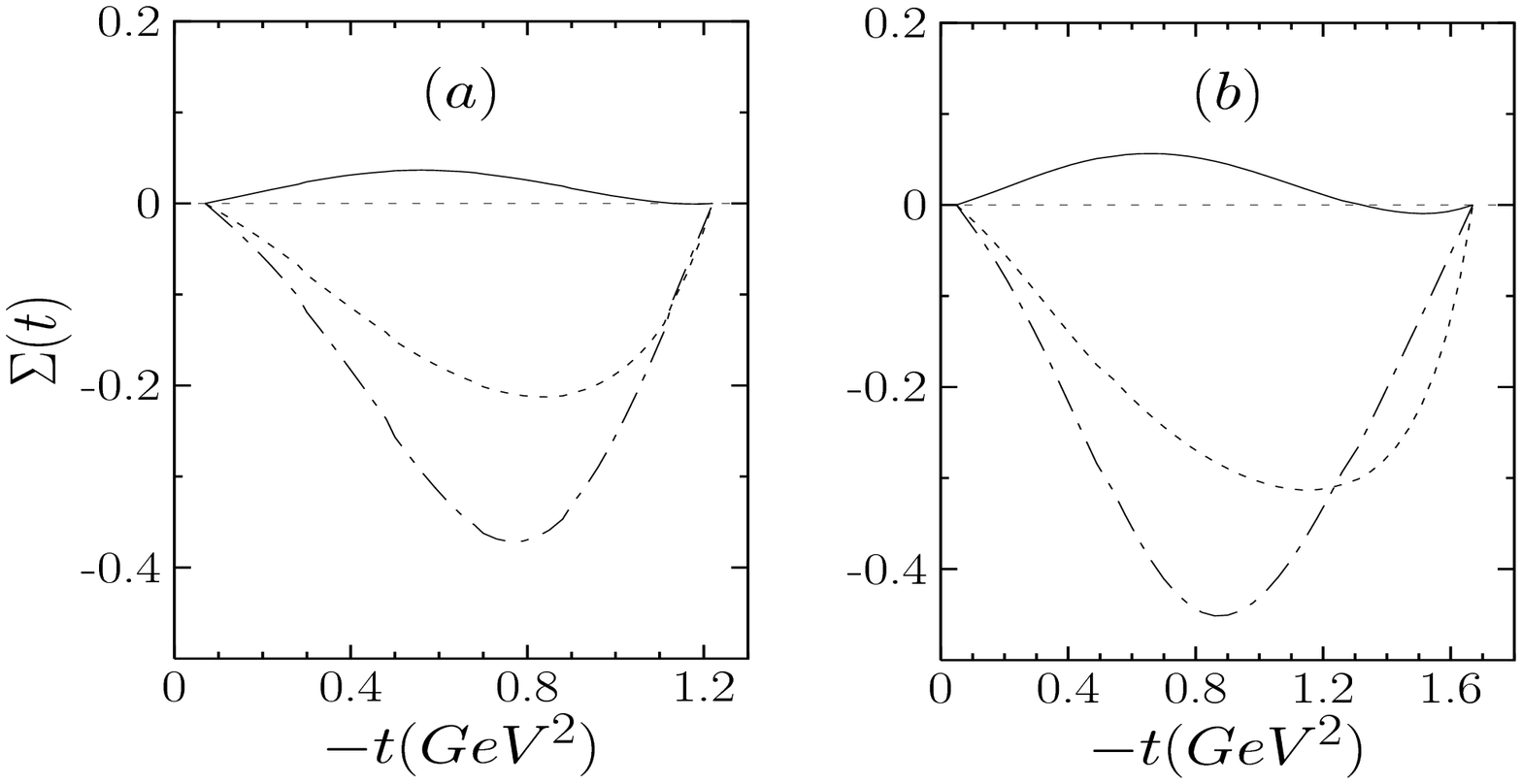}
\vskip 5.5cm
\caption{ Predicted behaviour of beam asymmetry for
$\gamma+p\rightarrow p+\omega$ at (a) $E_{\gamma}=$ $1.45$~GeV and
(b) $E_{\gamma}=$ $1.68$~GeV. Notation for different graphs is the same as
in Fig.2.}
\end{figure}
\newpage
\begin{figure}
$\left. \right.$
\vskip 10cm
    \includegraphics{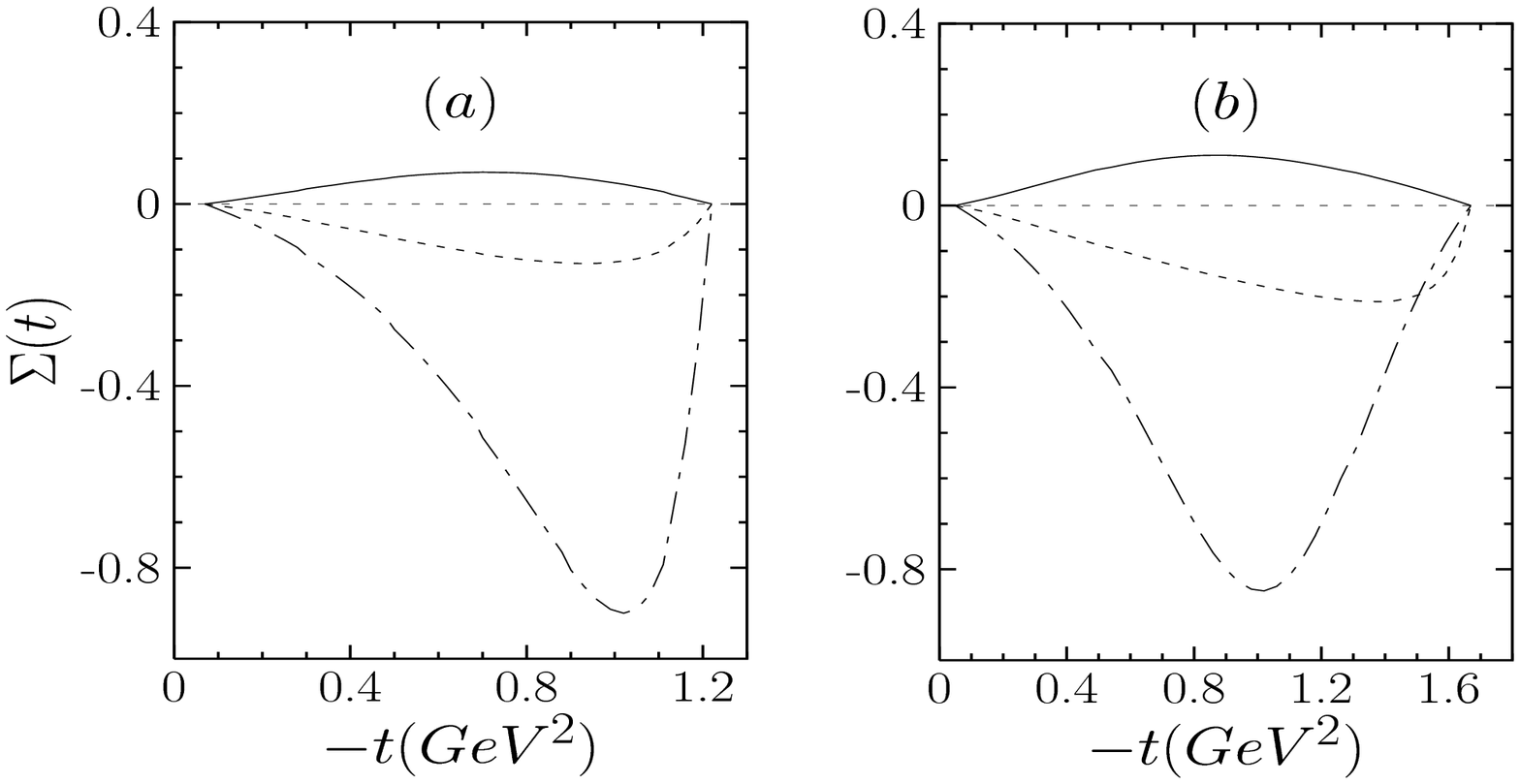}
\vskip 5.5cm
\caption{ Predicted behaviour of beam asymmetry for
$\gamma+n\rightarrow n+\omega$ at (a) $E_{\gamma}=$ $1.45$~GeV and
(b) $E_{\gamma}=$ $1.68$~GeV. Notation for different graphs is the same as
in Fig.2.}
\end{figure}
\newpage
\begin{figure}
$\left. \right.$
\vskip 10cm
    \includegraphics{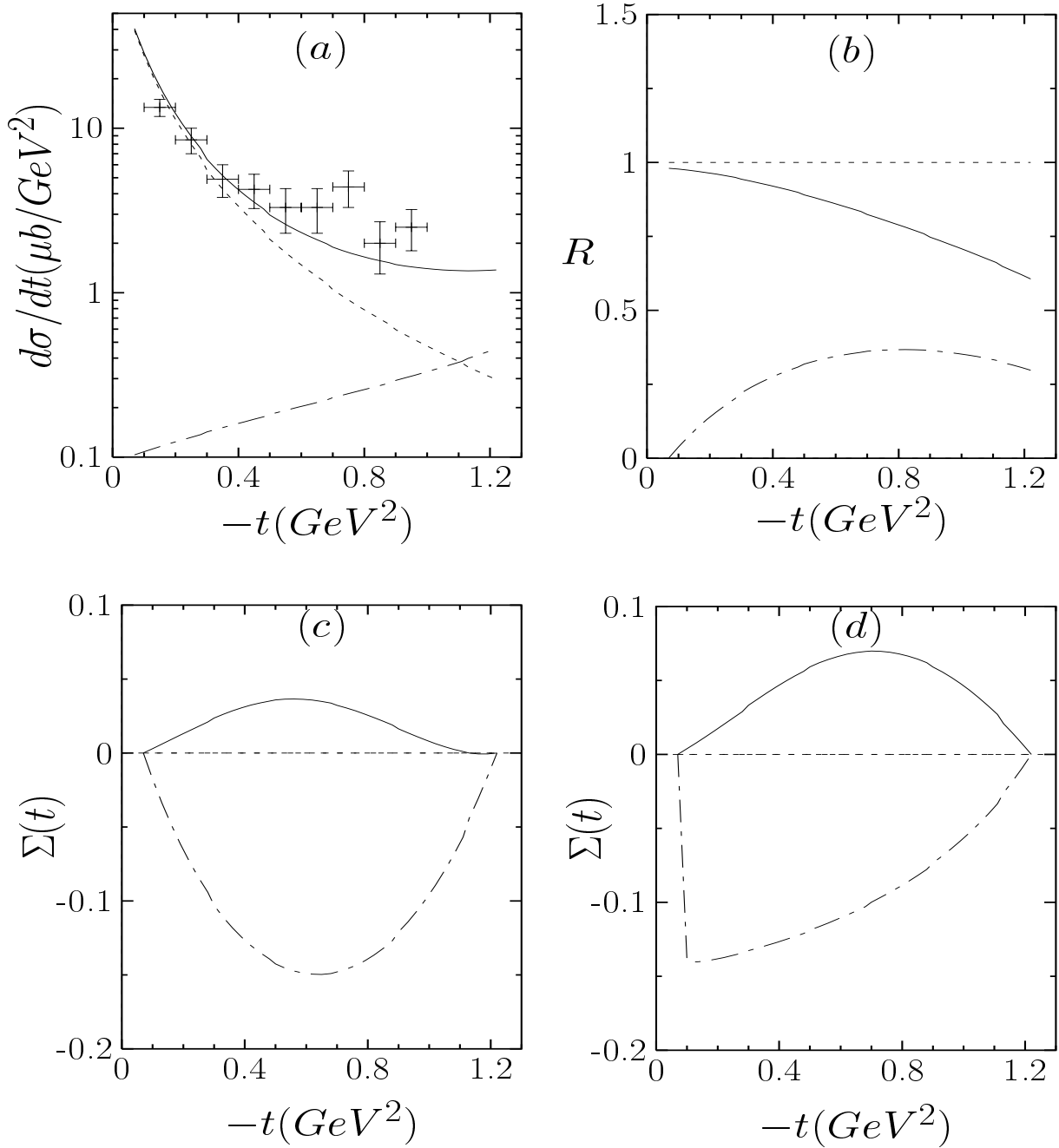}
%hscale=80 vscale=80 angle=0}
\vskip 8.5 cm
\caption{Different contributions of exchange mechanisms  to:
(a) $d\sigma(\gamma p\rightarrow
p\omega)/dt$, (b) $R=d\sigma(\gamma n\rightarrow n\omega)/d\sigma
(\gamma p\rightarrow p\omega)$, (c) $\Sigma(\gamma p\rightarrow p
\omega)$, (d) $\Sigma(\gamma n\rightarrow n\omega)$ at
$E_{\gamma}=1.45$~GeV for $g_{\omega NN}^{V}=0.5,~g_{\omega
NN}^{T}=0.1$ . Solid, dashed and dot-dashed lines correspond to
total, $\pi$-exhange and (s+u)-nucleon contributions,
respectively.}
\end{figure}
\end{document}